\RequirePackage[bookmarksnumbered,unicode]{hyperref}
\documentclass[sigconf]{acmart}

\usepackage{graphicx}
\usepackage{enumitem}
\setlist[itemize]{leftmargin=*}

\usepackage{hyperxmp}

\AtBeginDocument{%
  \providecommand\BibTeX{{%
    \normalfont B\kern-0.5em{\scshape i\kern-0.25em b}\kern-0.8em\TeX}}}

\begin{document}

\title{The Impact of Group Discussion and Formation on Student Performance: An Experience Report in a Large CS1 Course}

\author{Tong Wu}
\affiliation{%
  \institution{Virginia Tech}
  \city{Blacksburg}
  \state{VA}
  \country{USA}
}
\email{tongw@vt.edu}

\author{Xiaohang Tang}
\affiliation{%
  \institution{Virginia Tech}
  \city{Blacksburg}
  \state{VA}
  \country{USA}
}
\email{xiaohangtang@vt.edu}

\author{Sam Wong}
\affiliation{%
  \institution{University of Washington}
  \city{Seattle}
  \state{WA}
  \country{USA}
}
\email{samw627@uw.edu}

\author{Xi Chen}
\affiliation{%
  \institution{University of Virginia}
  \city{Charlottesville}
  \state{VA}
  \country{USA}
}
\email{xic@vt.edu}

\author{Clifford A. Shaffer}
\affiliation{%
  \institution{Virginia Tech}
  \city{Blacksburg}
  \state{VA}
  \country{USA}
}
\email{shaffer@cs.vt.edu}

\author{Yan Chen}
\affiliation{%
  \institution{Virginia Tech}
  \city{Blacksburg}
  \state{VA}
  \country{USA}
}
\email{ych@vt.edu}

\renewcommand{\shortauthors}{xx}

\begin{abstract}

Programming instructors often conduct collaborative learning activities, such as Peer Instruction (PI), to enhance student motivation, engagement, and learning gains. However, the impact of group discussion and formation mechanisms on student performance remains unclear. To investigate this, we conducted an 11-session experiment in a large, in-person CS1 course. We employed both random and expertise-balanced grouping methods to examine the efficacy of different group mechanisms and the impact of expert students' presence on collaborative learning. Our observations revealed complex dynamics within the collaborative learning environment. Among 255 groups, 146 actively engaged in discussions, with 96 of these groups demonstrating improvement for poor-performing students. Interestingly, our analysis revealed that different grouping methods (expertise-balanced or random) did not significantly influence discussion engagement or poor-performing students' improvement. In our deeper qualitative analysis, we found that struggling students often derived benefits from interactions with expert peers, but this positive effect was not consistent across all groups. We identified challenges that expert students face in peer instruction interactions, highlighting the complexity of leveraging expertise within group discussions.


\end{abstract}



\maketitle

\section{Introduction}
Collaborative learning has become an integral component of computer science education, particularly in introductory programming courses (CS1)~\cite{sabin1994collaborative}. This approach actively engages students in working together, learning new concepts, solving problems, and providing peer feedback. Programming instructors often employ various collaborative learning activities, such as Peer Instruction~\cite{10.1145/2445196.2445250Halvingfailrates}, pair programming~\cite{10.1145/3017680.3017748vPairProgramming}, and project-based work, to enhance students' motivation, engagement, and learning outcomes~\cite{serrano2014evaluation}. Among these, Peer Instruction (PI) has gained significant traction as an effective strategy where students individually respond to conceptual questions, discuss with peers, and then revise their answers, leading to improved failure rates, retention, and exam performance in Computer Science~\cite{taylor2018multi}. Several in-class coding exercise tools has developed to assist instructors in facilitating collaborative learning in programming classrooms. 
For example, PuzzleMe~\cite{wang2021puzzleme} enabled real-time peer discussions, balancing code and expertise diversity among students. 
VizPI~\cite{Tang2023VizPI} supports in-class Peer Instruction by allowing instructors to distribute coding exercises during lectures for synchronous student participation.
Große and Renkl's work has shown that grouping students with different solutions has been effective in improving learning outcomes in mathematics \cite{grosse2006effects}. However, the effectiveness of these discussions and different group formations on student coding performance remain areas of active research.

In this paper, we report our experience in conducting a series of collaborative learning sessions at our institute. Our analysis focuses on three key areas: 1) the effects of group discussions on students' coding performance, 2) the impact of different grouping mechanisms (random and expertise-balanced) on discussion activity and student coding performance, and 3) the influence of group composition (varying expertise levels) on discussion quality and student performance. This experiment involved a large-scale study, encompassing 788 student participants engaged in collaborative learning activities over 176 minutes across multiple sessions. The study generated extensive data, including interactions within 255 formed groups, analysis of group discussions, and evaluation of numerous coding submissions. 

Our experiment yielded several intriguing results that challenge common assumptions about collaborative learning in programming education. We found that active group participation did not consistently lead to improved coding performance for struggling students. The presence of expert students (those who achieved a 100\% pass rate on programming tasks prior to group formation), while often beneficial, did not guarantee positive outcomes for all group members. In fact, we observed instances where expert-led discussions failed to result in significant improvements for lower-performing students. Furthermore, our study revealed that the grouping method (expertise-balanced or random) had no significant impact on discussion activeness or overall student improvement across sessions. This finding suggests that the effectiveness of collaborative learning may depend more on the quality of interactions and individual student factors than on predetermined group composition strategies. Our experience also uncovered several challenges in implementing collaborative learning in large CS1 classes: the difficulty of achieving ideal expertise-balanced groups in real classroom settings, the tendency of some students to prioritize individual coding over group discussion, and the varying ability of expert students to effectively communicate with their peers. Interestingly, we noted that less active participants sometimes benefited from "lurking" in group discussions, suggesting that silent participation can also be a valid form of learning in these environments. These insights provide valuable lessons for future research on grouping mechanisms and in-class activity design. 


\section{Related Work}

Collaborative learning in computer science education has demonstrated promising results in improving student engagement, performance, and retention. Studies have shown increased student confidence, participation, and proficiency in introductory programming courses ~\cite{falkner2009easing}. Programming instructors often employ various collaborative learning activities, such as peer instruction, pair programming, and project-based work, to enhance students' motivation, engagement, and learning outcomes. Zingaro and Porter ~\cite{zingaro2014peer} investigated the impact of peer discussions on student performance in computer science courses and found that such discussions can lead to significant learning gains, with instructor-led discussions being particularly valuable for all student groups. However, collaborative learning is not without challenges. James and Willoughby ~\cite{james2011listening} observed that over one-third of in-class peer discussions in large classes can be unproductive, highlighting the need for careful design of collaborative activities and group formation.

Group formation in computer science classes is crucial for promoting effective collaborative learning. Deibel ~\cite{deibel2005team} found that instructor-selected teams for in-class group work can enhance student interaction and learning. Automated group formation mechanisms have shown promise in creating consistent and successful groups for learning activities ~\cite{webber2012evaluating}. However, forming productive groups remains challenging due to the complexity of the problem and the sparsity of the solution space ~\cite{henry2013forming}. To address this, researchers have proposed various approaches, including grouping by learning style, latent jigsaw methods ~\cite{deibel2005team}, and massively parallel brute-force algorithms ~\cite{henry2013forming}. Web-based applications that collect student information and use algorithms to form groups offer a more systematic approach to group formation ~\cite{henry2013forming}. Nevertheless, the effectiveness of collaborative activities in technical domains such as logic programming remains uncertain.

To support collaborative programming, researchers have developed various tools and platforms. VizGroup facilitates visual collaboration for distributed software teams~\cite{tang2024vizgroup}. Edcode enables real-time collaborative coding and automated assessment~\cite{chen2020edcode}. Lee et al. explored coordination models for ad hoc programming teams~\cite{lee2017exploring}. Wang et al. ~\cite{wang2021puzzleme} developed PuzzleMe, a tool that enables real-time peer discussions for programming learners while balancing code and expertise diversity. These tools have facilitated the distribution of exercises, grouping of students into small teams, and real-time monitoring of student performance.

\section{Experiment Setup}

\subsection{CS1 Class}
Our experiment was conducted at the author's institution. We chose the Introduction to Programming in Python course during the Spring 2024 semester. With a total enrollment of 694 students spread across three sessions, this course is co-taught by two instructors and supported by a team of 21 TAs. The course is structured entirely as an in-person experience. The curriculum covers a wide range of fundamental programming concepts, including basic control flow with loops and conditionals, state tracing and manipulation, simple and complex data types, functional and object-oriented coding strategies, and data processing. Our experiment was conducted across 11 sessions as detailed in Table ~\ref{session_data}, leveraging the teaching schedules of both instructors. The course's structure allowed us to seamlessly implement our group discussion and peer instruction activities within the existing framework, which already included regular in-class activities, allowed us to seamlessly implement our group discussion and peer instruction activities. 

\begin{figure}[t]
\centering
\includegraphics[width=1\linewidth]{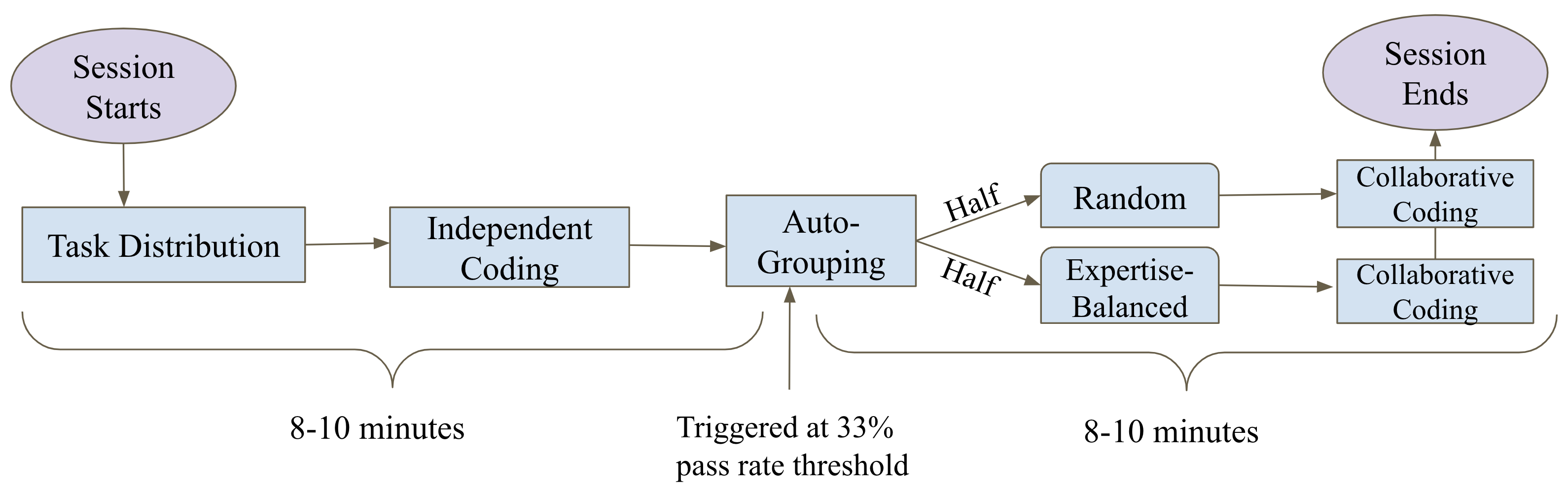}
\caption{High-level experimental workflow}
\label{workflow}
\end{figure}
\subsection{In-class Exercises}
Figure~\ref{workflow} illustrates the procedure of the in-class experiments. All experiments were conducted in the CS1 course. Before starting each experiment, an instructor announced the procedure of the peer instruction session and distributed the programming exercise to all students. Students then worked independently on the exercise until the overall class performance reached a $33\%$ pass rate. At this point, students were automatically divided into groups using two different mechanisms: half were grouped randomly, and the other half were grouped based on an expertise-balanced mechanism (see Sec~\ref{sec: group conditions}). During the collaborative coding phase, students in each group shared a text-based chat channel where they could see each other's pass rates in real time. The session concluded when the instructor manually terminated the chat channels. Both the independent and collaborative coding parts lasted approximately 8-10 minutes each. We collected students' chat and code submission data during the experiment sessions, ensuring that no identifying information was included.

All the programming tasks in the experiment sessions were designed by the instructors of the course, matching students' knowledge levels and the course progress. Several learning objectives such as \textit{Variables}, \textit{Functions}, \textit{Strings}, \textit{Lists}, and \textit{Statements (e.g., if, for, while)} were covered in the tasks.

\subsection{Data Collection System}

\begin{figure}[]
    \centering
\includegraphics[width=1\linewidth]{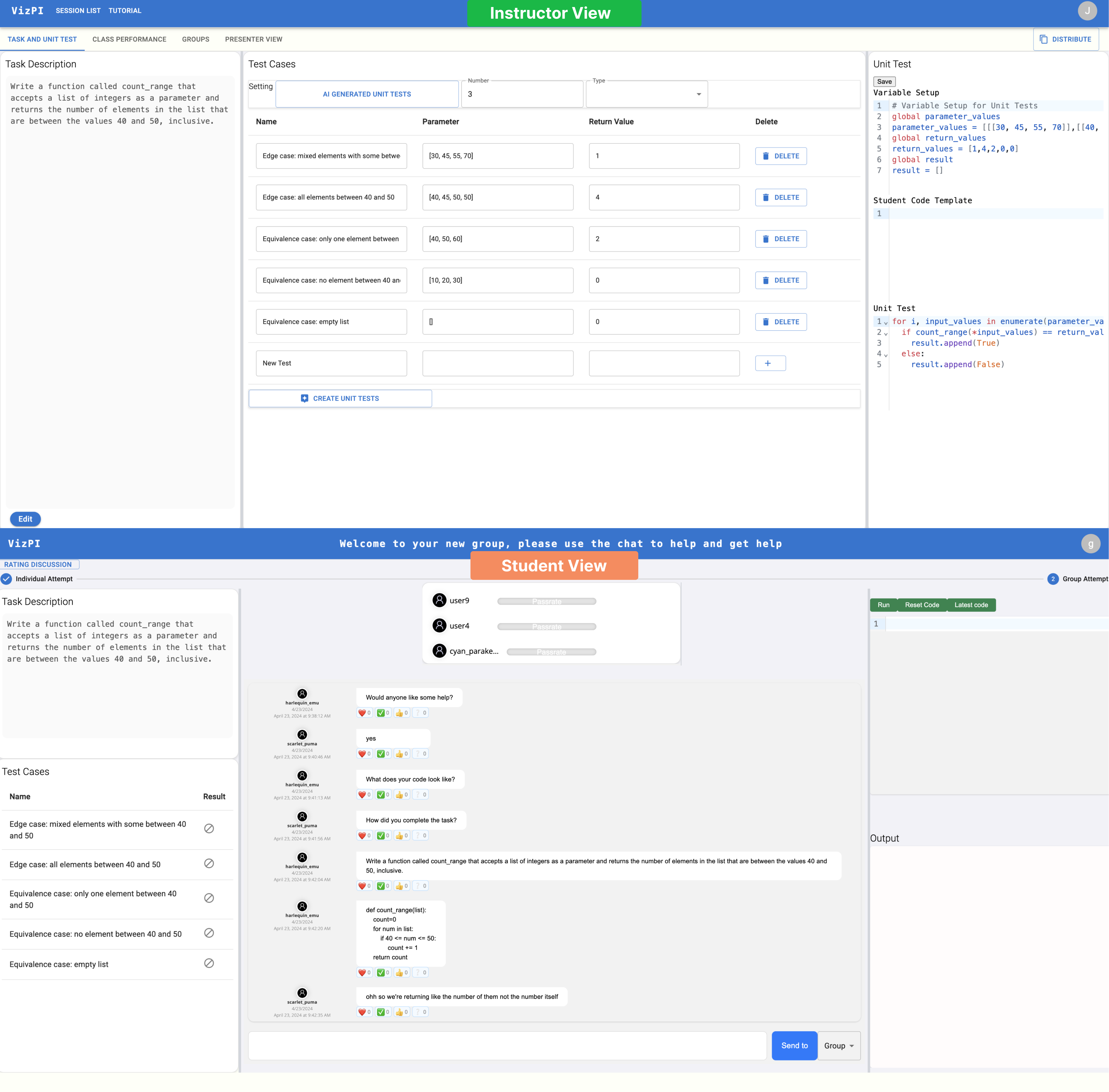}
    \caption{Instructor and student UI of VizPI}
    \label{UI}
\end{figure}

We used a learning platform VizPI~\cite{Tang2023VizPI} to collect student discussion and coding data. The interface (see Figure~\ref{UI}) consists of an instructor UI and a student UI. The instructor UI enables instructors to easily conduct peer instruction activity by creating and distributing coding exercises and dividing the class into small groups. The student UI features an IDE and group chat window where students can write and test their code while communicating with their peers. In each group, students remain anonymous, and they can view each other's pass rate but not their code.

\subsection{Grouping Mechanisms}
\label{sec: group conditions}
The system first collects comprehensive data on all participating users, focusing on the number of messages sent and their code acceptance rates. Using the Zigzag Distribution Algorithm ~\cite{tran2004peer}, the system then divides the students into two groups:
\begin{itemize}
    \item \textbf {Random Group:} Students are randomly assigned into groups of three or four.
    \item \textbf {Expertise-balanced Group:} Students are categorized into high, medium, and low performers based on their pass rate.
\end{itemize}
The system further analyzes the similarity of students' code submissions using the text-embedding-3-large model\footnote{\url{https://platform.openai.com/docs/models/embeddings}}to generate embeddings, then compares these embeddings using cosine similarity. To maximize within-group code similarity, each group is intentionally formed with one student from each performance category (high, medium, low), ensuring a balanced mix of skills and perspectives. Any remaining students are grouped to ensure no one is left out.

\section{Results \& lessons learned}

\begin{table*}[ht]
\centering
\resizebox{\textwidth}{!}{%
\begin{tabular}{|c|c|c|c|c|c|c|c|c|c|c|}
\hline
\textbf{Session ID} & \textbf{\begin{tabular}[c]{@{}c@{}}Student total\\ number\end{tabular}} & \textbf{\begin{tabular}[c]{@{}c@{}}Balanced group\\ number\end{tabular}} & \textbf{\begin{tabular}[c]{@{}c@{}}Students in\\ balanced group\end{tabular}} & \textbf{\begin{tabular}[c]{@{}c@{}}Random group\\ number\end{tabular}} & \textbf{\begin{tabular}[c]{@{}c@{}}Students in\\ random group\end{tabular}} & \textbf{Session start time} & \textbf{Grouping time} & \textbf{Session end time} & \textbf{\begin{tabular}[c]{@{}c@{}}Session lasting\\  time\end{tabular}} & \textbf{\begin{tabular}[c]{@{}c@{}}Discussion lasting\\ time\end{tabular}} \\
\hline
1 & 52 & 8 & 26 & 8 & 26 & 2024-04-22 09:02:42 & 09:15:47 & 09:22:46 & 0:20:04 & 0:06:59 \\ \hline
2 & 58 & 9 & 29 & 9 & 29 & 2024-04-22 09:23:20 & 09:42:16 & 09:45:11 & 0:21:51 & 0:02:55 \\ \hline
3 & 78 & 13 & 39 & 13 & 39 & 2024-04-22 12:32:27 & 12:34:20 & 12:39:53 & 0:07:26 & 0:05:33 \\ \hline
4 & 84 & 14 & 42 & 14 & 42 & 2024-04-22 12:46:07 & 12:50:16 & 12:53:59 & 0:07:52 & 0:03:43 \\ \hline
5 & 82 & 13 & 41 & 13 & 41 & 2024-04-23 09:34:54 & 09:38:13 & 09:51:45 & 0:16:51 & 0:13:32 \\ \hline
6 & 81 & 13 & 41 & 13 & 40 & 2024-04-24 12:26:32 & 12:34:16 & 12:37:58 & 0:11:26 & 0:03:42 \\ \hline
7 & 62 & 10 & 31 & 10 & 31 & 2024-04-24 09:26:54 & 09:40:01 & 09:43:02 & 0:16:08 & 0:03:01 \\ \hline
8 & 63 & 10 & 32 & 10 & 31 & 2024-04-24 09:39:45 & 09:50:53 & 09:55:42 & 0:15:57 & 0:04:49 \\ \hline
9 & 67 & 11 & 34 & 11 & 33 & 2024-04-24 09:00:52 & 09:16:31 & 09:30:42 & 0:29:50 & 0:14:11 \\ \hline
10 & 83 & 14 & 42 & 13 & 41 & 2024-04-24 12:46:12 & 12:49:59 & 13:00:05 & 0:13:53 & 0:10:06 \\ \hline
11 & 78 & 13 & 39 & 13 & 39 & 2024-04-24 13:00:59 & 13:02:59 & 13:15:50 & 0:14:51 & 0:12:51 \\
\hline
\end{tabular}%
}
\caption{Session's Student Number, Grouping Number, and Time Frame Data}
\label{session_data}
\end{table*}
\subsection{Basic Stats \& Data Cleaning}

We collected data from 11 sessions of CS1 Class, as detailed in Table~\ref{session_data}. These sessions involved a total of 862 students, of which 788 students fully participated and 74 were excluded due to late arrivals or being ungrouped. Individual session participation ranged from 52 to 84 students. We formed 255 groups, split evenly between expertise-balanced (128 groups) and random assignment (127 groups) mechanisms. Within both grouping mechanisms, we further categorized groups based on their composition, focusing primarily on the presence of expert students (those achieving a 100\% pass rate before grouping). Our improvement analysis focused on these poor-performing students (Pass Rate < 100\%). The experiment generated rich interaction data:
\begin{itemize}
    \item 598 messages exchanged among participants
    \item Session durations ranging from 7:26 to 29:50 minutes (average $\sim$16 minutes)
    \item Discussion phases lasting 2:55 to 14:11 minutes (average $\sim$7 minutes)
\end{itemize}

Our data cleaning process involved filtering messages and normalizing time data to ensure comparability across sessions. This comprehensive dataset and our categorization of group expertise levels provide a foundation for analyzing group discussion effectiveness and the impact of different grouping mechanisms on student performance.

\subsection{Group and Discussion Definitions}
\begin{figure}[h]
\centering
\includegraphics[width=1\linewidth]{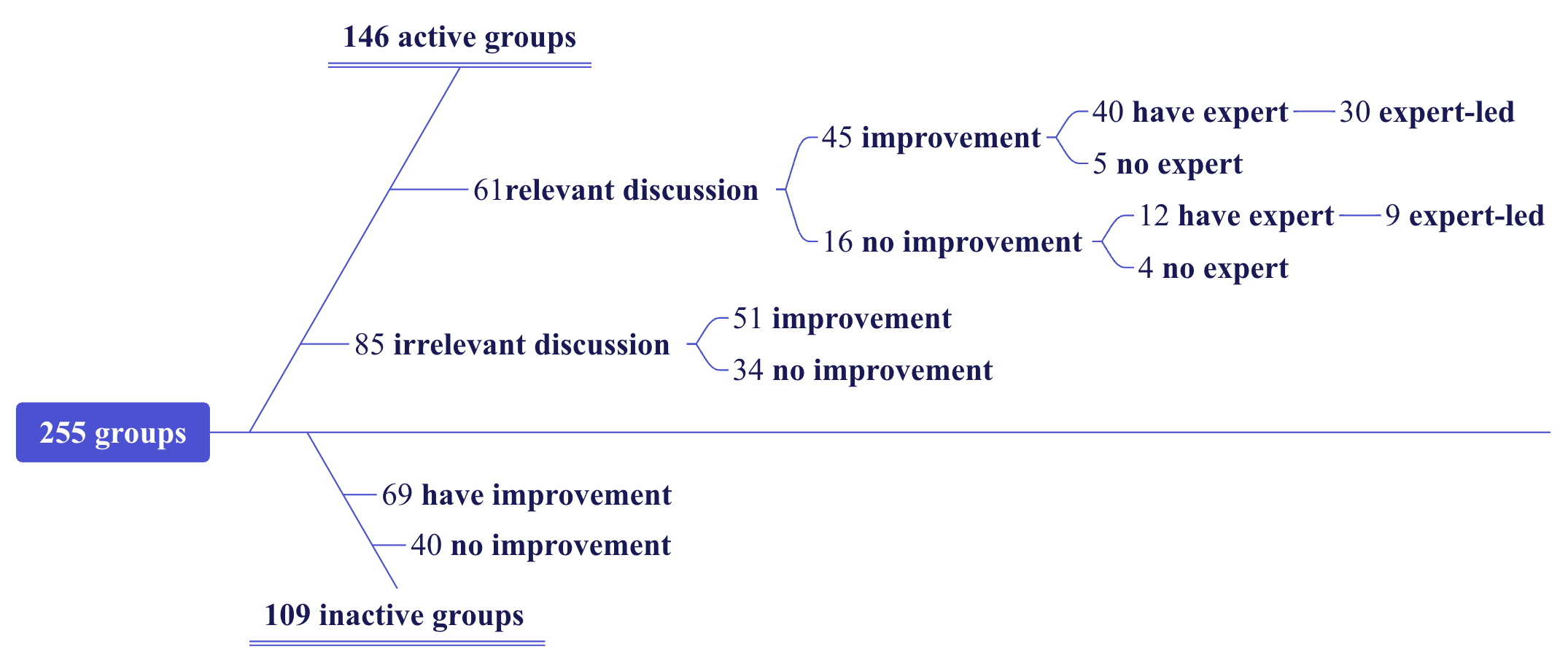}
\caption{Groups activity and improvement}
\label{255 groups}
\end{figure}

We established several key definitions and categorized groups accordingly to analyze group dynamics and their impact on student coding performance. 
\begin{flushleft} 
\small
\begin{itemize}
    \item \textbf{Active groups} Groups where messages were exchanged
    \item \textbf{Inactive groups} Groups with no message exchanges
    \item \textbf{Relevant discussion} Instances where at least two students exchanged messages discussing code issues or exercise approaches
    \item \textbf{Irrelevant discussion} Groups with only one message, or where messages were limited to default greetings without substantive replies about the coding task
    \item \textbf{Improvement} Poor-performing students increased their pass rate from pre-grouping to the end of the session
    \item \textbf{No improvement} No increase in pass rate for poor-performing students
    \item \textbf{Have expert} Groups with at least one expert student
    \item \textbf{No expert} Groups without any expert students
    \item \textbf{Expert-led discussion} Discussions led by expert students
\end{itemize}
\end{flushleft}
Figure~\ref{255 groups} illustrates the breakdown of our 255 groups based on these definitions.


\subsection{Impact of Active and Relevant Group Discussions on Students' Coding Performance}

 Our analysis revealed that active and relevant discussions generally correlate with improved coding performance, but this relationship is not consistent across all sessions. The effectiveness of group discussions varies widely, with improvement rates ranging from 0\% to 85\% across sessions. Interestingly, we observed that inactive or irrelevant discussions sometimes led to better performance, highlighting the complexity of collaborative learning in programming education.

\subsubsection{Active vs. Inactive Groups}

\begin{figure}[h]
    \centering
\includegraphics[width=1\linewidth]{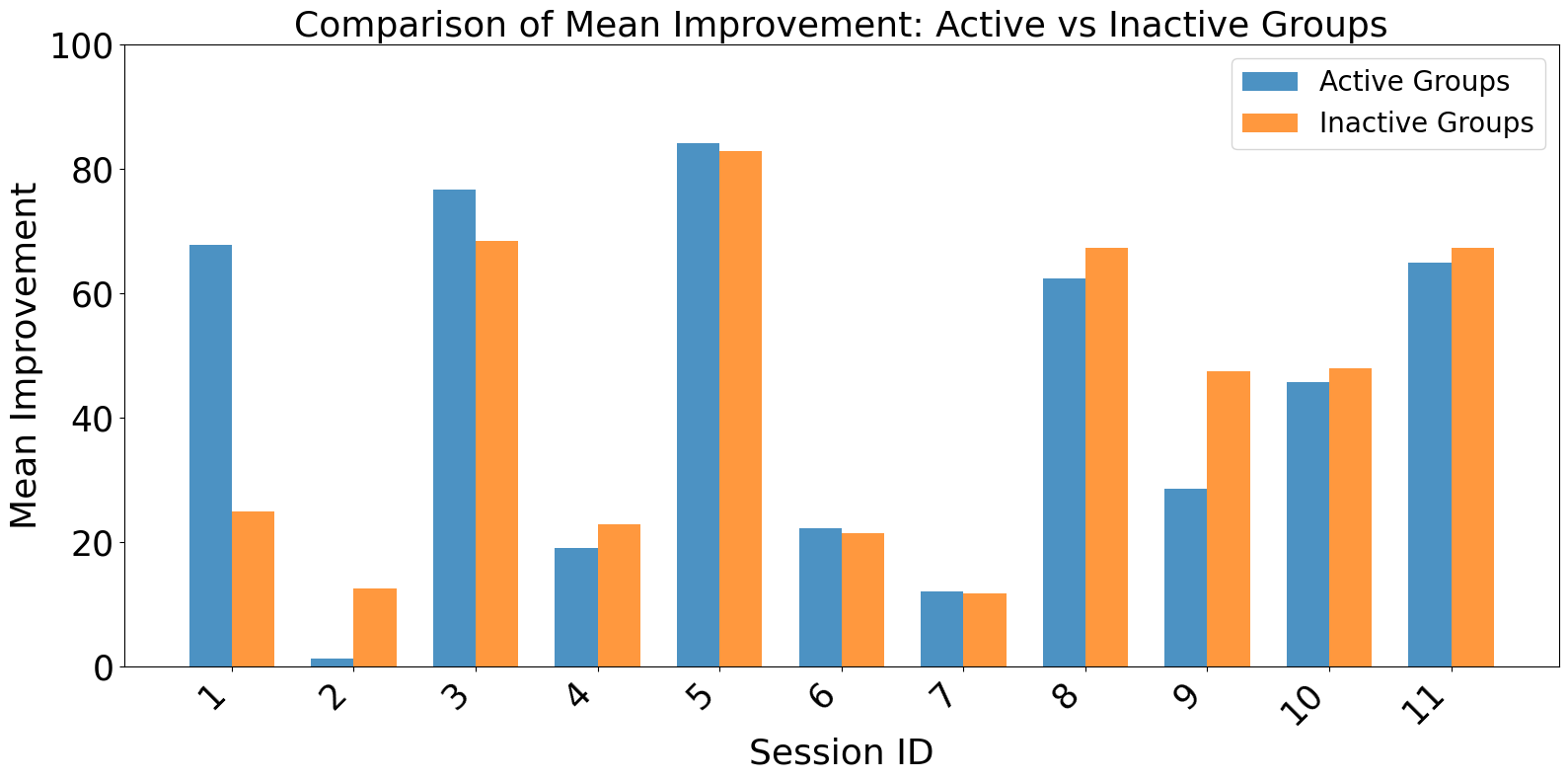}
    \caption{Active and improvement}
    \label{Active and improvement}
\end{figure}

Figure ~\ref{Active and improvement} illustrates the breakdown of active and inactive groups and their impact on student improvement. Among the active groups, 96 (66\%) showed improvement in poor-performing students' coding skills, compared to 69 (63\%) in inactive groups. However, the session-by-session analysis revealed inconsistencies:

\begin{itemize}
    \item In 7 out of 11 sessions (64\%), active groups showed higher mean improvement than inactive groups.
    \item In 4 sessions (36\%), inactive groups outperformed active groups.
    \item The magnitude of difference between active and inactive group performance varied widely across sessions.
\end{itemize}

\subsubsection{Relevant vs. Irrelevant Discussions}

\begin{figure}[h]
    \centering
\includegraphics[width=1\linewidth]{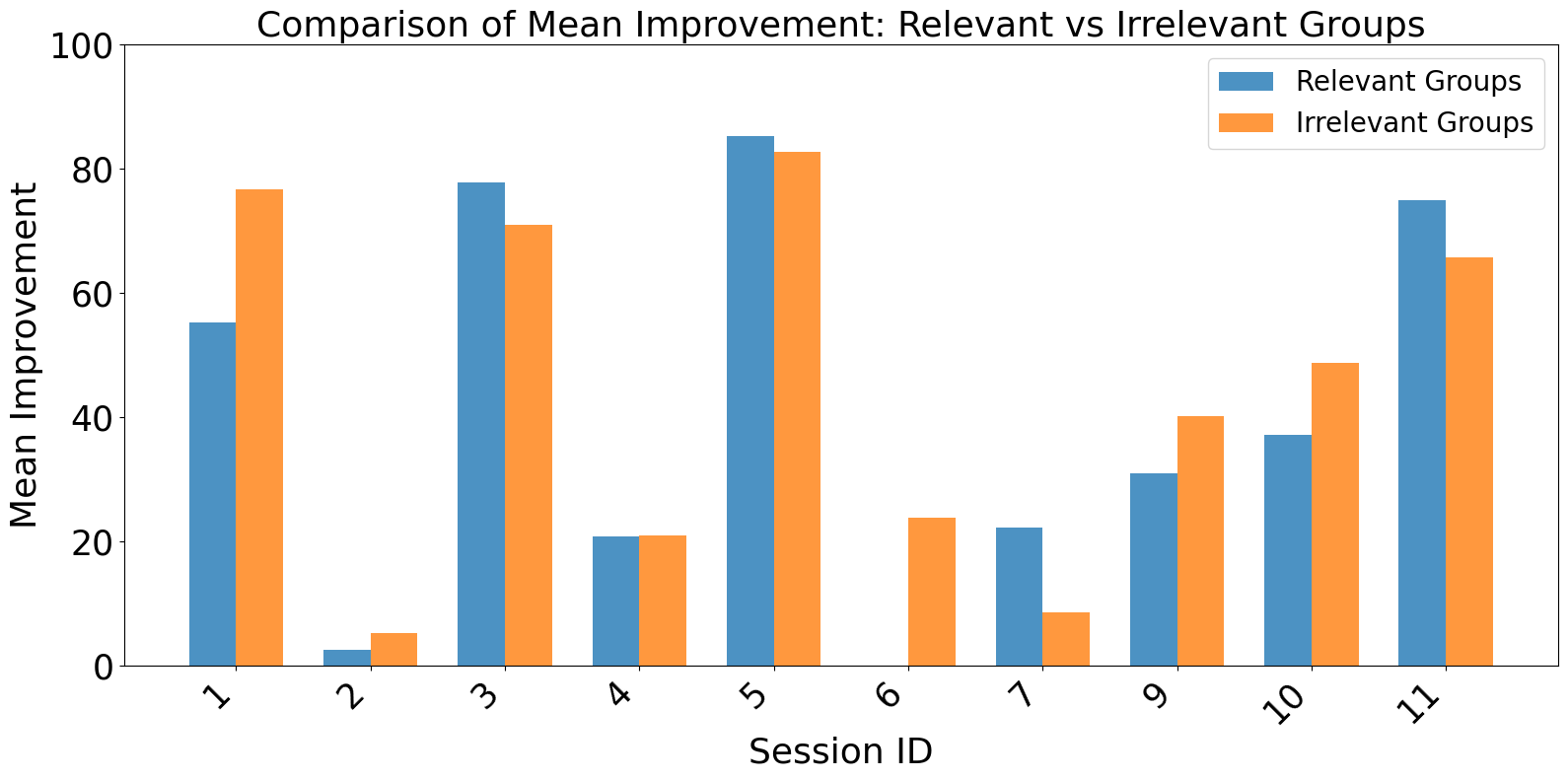}
    \caption{Relevant and improvement}
    \label{Relevant and improvement}
\end{figure}

To explore the quality of discussions, we examined the relevance of discussions within active groups (see Figure ~\ref{Relevant and improvement}). Of the groups with relevant discussions, 45 (74\%) showed improvement, compared to 51 (60\%) of groups with irrelevant discussions. This analysis provided additional insights:

\begin{itemize}
    \item In 6 out of 10 sessions (60\%), groups with relevant discussions showed higher mean improvement than those with irrelevant discussions. (Session6 has no relevant discussions)
    \item In 4 sessions (40\%), groups with irrelevant discussions unexpectedly outperformed those with relevant discussions.
    \item The impact of discussion relevance on performance improvement varied considerably across sessions.
\end{itemize}

\subsubsection{Implications of Findings and Lessons Learned}
Our observations suggest that the effectiveness of group discussions may be influenced by multiple factors beyond mere participation or perceived relevance. While not conclusively proven in this study, factors such as task difficulty, student expertise levels, and other contextual elements likely play significant roles. Further research is needed to quantify these effects. Besides, the wide variation in improvement rates across sessions underscores the highly context-dependent nature of group discussions' impact. This observation teaches us that a one-size-fits-all approach to collaborative learning in programming education is insufficient. Future experiments should incorporate more detailed contextual data to better understand these variations.\\
Perhaps most notably, the performance of some inactive or irrelevant discussion groups highlights the multifaceted nature of learning in programming education. Their performance proves students' independent trial-error process could lead to performance improvement. This finding suggests that our current metrics for categorizing discussions as "active" or "relevant" may not fully capture the nuanced ways in which students learn from peer interactions. It also indicates that silent participation or even seemingly off-topic discussions might have unexpected benefits that our current measurement tools fail to capture. 

\subsection{The effectiveness of different grouping methods (random vs. expertise-balanced) on discussion activity and student coding performance}

\begin{figure}[]
    \centering
\includegraphics[width=1\linewidth]{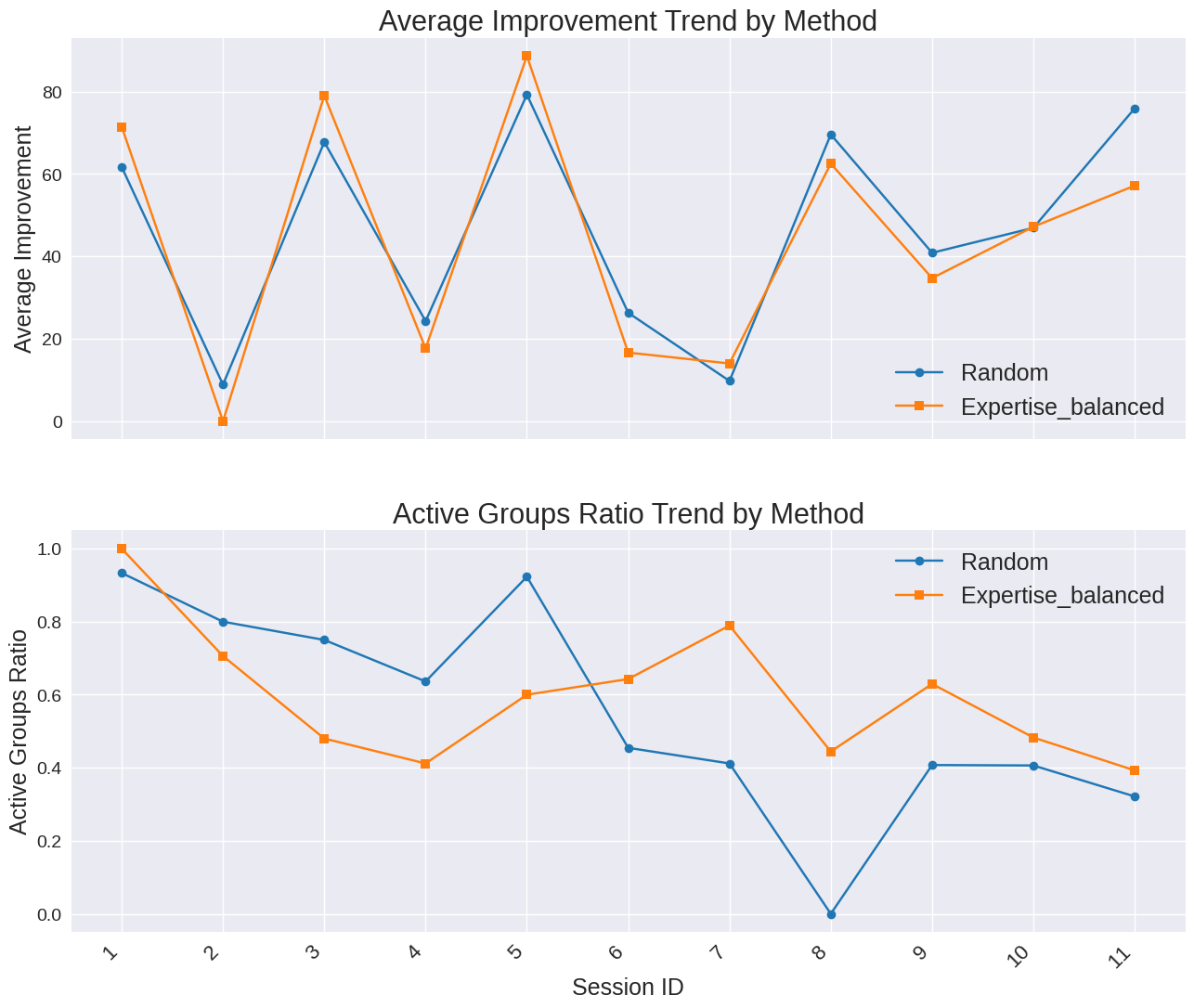}
    \caption{Comparison of Random and Expertise-balanced Group Formation Methods Across Multiple Sessions}
    \label{group type}
\end{figure}

Our experiment compared two grouping mechanisms: random and expertise-balanced grouping. We analyzed their impact on group discussion activity and student coding performance across 11 sessions (see Figure ~\ref{group type}). The results reveal that there’s no clear superiority of one method over the other.

\subsubsection{Discussion Activity}

The data shows considerable variation in active group ratios (the proportion of groups that engaged in discussion during a session) between sessions and grouping methods:

\begin{itemize}
    \item Random grouping: Active group ratios ranged from 0\% to 93\%, with a mean of 55\%.
    \item Expertise-balanced grouping: Active group ratios ranged from 41\% to 100\%, with a mean of 60\%.
\end{itemize}

While the expertise-balanced groups showed slightly higher average active group ratios, the difference was not consistent across all sessions. Some sessions (e.g., 1, 5) saw higher activity in expertise-balanced groups, while others (e.g., 2, 4) showed higher activity in randomly formed groups.

\subsubsection{Student Coding Performance}

We measured student coding performance improvement for both grouping methods:

\begin{itemize}
    \item Random grouping: Average improvement ranged from 9\% to 79\%, with an overall mean of 46\%.
    \item Expertise-balanced grouping: Average improvement ranged from 0\% to 89\%, with an overall mean of 44\%.
\end{itemize}

The overall average improvement was slightly higher for random grouping (46\%) compared to expertise-balanced grouping (44\%). However, this difference was not statistically significant, as indicated by the Wilcoxon signed-rank test (statistic = 29.00, p-value = 0.7646). This suggests that the grouping method (random vs. expertise-balanced) does not significantly impact student improvement.

\subsubsection{Session-by-Session Analysis}

The effectiveness of each grouping method varied considerably across sessions:

\begin{itemize}
    \item In 6 out of 11 sessions, random grouping showed higher average improvement.
    \item In 5 out of 11 sessions, expertise-balanced grouping showed higher average improvement.
    \item The magnitude of difference in improvement between the two methods ranged from 0.2\% (session 10) to 19\% (session 11).
\end{itemize}

Notably, some sessions (e.g., 2) showed a stark contrast, with random grouping yielding an 9\% improvement while expertise-balanced grouping resulted in no improvement.

\subsubsection{Implications and Lessons Learned}

Our findings on the effectiveness of random versus expertise-balanced grouping mechanisms reveal a more nuanced picture than initially anticipated. The lack of a clear, consistent advantage for either method across all sessions highlights the complexity of collaborative learning in programming education. This complexity suggests that the effectiveness of grouping strategies may be influenced by a multitude of factors beyond the simple dichotomy of random versus balanced expertise. The varying degrees of success for both methods across different sessions indicate that contextual elements such as task difficulty, student preparedness, and even subtle variations in group dynamics play crucial roles in determining outcomes.

These observations lead us to consider that the apparent similarity in performance between random and expertise-balanced groups might be due to underlying similarities in group composition that occur naturally even in random grouping. With 61 out of 127 random groups having similar expertise levels to the expertise-balanced groups, we must question whether our current approach to expertise-balanced grouping is sufficiently distinct from random assignment to produce measurable differences in outcomes. This realization points to the need for a more granular analysis of group composition and its impact on both group activity and student performance.

\subsection{The influence of group composition (varying expertise levels) on discussion activity and student coding performance}

In this section, we delve deeper into the detailed group composition to examine whether the lack of significant differences between random and expertise-balanced groups is due to their similar expertise levels. While quantitative analysis across sessions did not yield consistent or statistically significant results, a deeper semantic analysis of group interactions provided valuable insights into the role of expertise in group discussions and its influence on student performance.

\subsubsection{Expert-led Discussions and Performance Improvement}

A notable trend emerged from our analysis of active discussion groups where poor-performing students showed improvement. Out of 40 such groups, 30 (75\%) were expert-led, meaning they included at least one expert student who actively participated in the discussion. This suggests that the presence of an expert can often catalyze productive discussions and lead to performance improvements for struggling students.

For instance, in group 2889 from session 3, we observed a beneficial interaction between an expert student (Student A) and a struggling student (Student B):

\begin{quote}
\small
Student A: Would anyone like some help?\\
Student B: Yes, I tried to do a list comprehension like this "def good\_scores(scores): new\_scores = [score for scores in good\_scores if score > 80] return new\_scores" should I do a for loop instead?\\
Student A: I used a for loop, but a comprehension should work. You need to change good\_scores to just scores since scores is the list.\\
Student B: ohh i fixed it\\
Student A: nice :)\\
Student B: thanks pink\_primate!
\end{quote}

This exchange demonstrates how an expert's presence can facilitate a focused, problem-solving dialogue, leading to immediate improvements in the struggling student's code.

\subsubsection{Limitations and Factors Influencing the Effectiveness of Expert Presence}
Our analysis also revealed that the mere presence of expert students does not guarantee successful knowledge transfer or performance improvement for struggling peers. Here is an example between two expert students (C and E) and one struggling student (Student D) from group 2860 in session 2:

\begin{quote}
\small
Student C: For those that didn't get it, please send in a copy of your code\\
Student D: "def remove\_vowels(phrase): vowels = ('A', 'E', 'I', 'O', 'U') for letter in phrase: if letter in vowels.upper(): phrase.strip(letter) elif letter in vowels.lower(): phrase.strip(letter) return phrase"\\
Student E: What errors are you getting?\\
Student C: I don't think you can strip a letter from the mddle of the word, only the ends\\
Student D: tuple' object has to attribute 'upper'\\
Student C: Write your values as a list and not as a tuple\\
Student C: Like this maybe: vowels = 'aeiouAEIOU'\\
Student E: that's a string, a list is contained in []
\end{quote}

In this instance, the struggling student failed to improve despite iterative attempts and group support. Our analysis across 40 groups identified several factors contributing to the varying effectiveness of these interactions: insufficient problem identification, fragmented or ambiguous advice, temporal constraints, conflicting guidance from multiple sources, and the absence of structured, step-by-step problem-solving methodologies. These factors were observed repeatedly in discussions where expert-led interactions failed to yield significant progress for struggling students.

These findings underscore the challenges faced by expert students when attempting to contribute effectively to group discussions. The efficacy of expert-led discussions is contingent upon the expert's ability to communicate clearly, identify core issues, and provide structured, tailored guidance to struggling peers. This observation elucidates why expertise-balanced group mechanisms do not consistently yield significant improvements in student performance. The key inference is that leveraging expertise within group discussions is a multifaceted challenge, emphasizing the necessity for targeted strategies to enhance peer instruction in collaborative coding environments. Future research should focus on developing and evaluating such strategies to optimize the benefits of Peer Instruction learning in computer science education.

















\section{Discussion \& Conclusion}
Our main observations from this experiment including: 1) Active discussions generally correlated with improved coding performance, but this relationship was not consistent across all sessions. The effectiveness of group discussions varied widely, with improvement rates ranging from 0\% to 85.22\% across sessions. 2) Different grouping method (expertise-balanced or random) does not significantly influence group discussion activeness and students' improvement. And 3) 75\% of productive groups (groups have active discussion and improved performance) were led by expert students. This finding aligns with Vygotsky's concept of the Zone of Proximal Development ~\cite{vygotsky1978mind}. Expert students often provided scaffolding for their struggling peers, facilitating knowledge transfer and problem-solving skills. However, the mere presence of an expert did not guarantee successful outcomes. Factors such as communication clarity, problem-identification skills, and the ability to provide structured guidance influenced the effectiveness of expert-led discussions. This complexity echoes findings by James and Willoughby~\cite{james2011listening}, who observed that a significant portion of in-class peer discussions can be unproductive.

Our experiment highlighted several challenges in implementing effective collaborative learning strategies in large CS1 courses. Firstly, achieving an ideal expertise balance in real classroom settings proved to be a complex task, often deviating from theoretical expectations. Secondly, we observed a tendency among some students to prioritize individual coding efforts over engaging in group discussions, potentially limiting the benefits of collaborative learning. Lastly, the effectiveness of grouping mechanisms, whether random or expertise-balanced, exhibited considerable variation across sessions. This variability suggests that uniform approaches to group formation may be insufficient in addressing the diverse learning needs and dynamics present in large-scale programming courses.

\bibliographystyle{ACM-Reference-Format}
\bibliography{software.bib}


\begin{thebibliography}{19}


\ifx \showCODEN    \undefined \def \showCODEN     #1{\unskip}     \fi
\ifx \showDOI      \undefined \def \showDOI       #1{#1}\fi
\ifx \showISBNx    \undefined \def \showISBNx     #1{\unskip}     \fi
\ifx \showISBNxiii \undefined \def \showISBNxiii  #1{\unskip}     \fi
\ifx \showISSN     \undefined \def \showISSN      #1{\unskip}     \fi
\ifx \showLCCN     \undefined \def \showLCCN      #1{\unskip}     \fi
\ifx \shownote     \undefined \def \shownote      #1{#1}          \fi
\ifx \showarticletitle \undefined \def \showarticletitle #1{#1}   \fi
\ifx \showURL      \undefined \def \showURL       {\relax}        \fi
\providecommand\bibfield[2]{#2}
\providecommand\bibinfo[2]{#2}
\providecommand\natexlab[1]{#1}
\providecommand\showeprint[2][]{arXiv:#2}

\bibitem[Chen et~al\mbox{.}(2020)]%
        {chen2020edcode}
\bibfield{author}{\bibinfo{person}{Yan Chen}, \bibinfo{person}{Jaylin Herskovitz}, \bibinfo{person}{Gabriel Matute}, \bibinfo{person}{April Wang}, \bibinfo{person}{Sang~Won Lee}, \bibinfo{person}{Walter~S Lasecki}, {and} \bibinfo{person}{Steve Oney}.} \bibinfo{year}{2020}\natexlab{}.
\newblock \showarticletitle{EdCode: Towards Personalized Support at Scale for Remote Assistance in CS Education}. In \bibinfo{booktitle}{\emph{2020 IEEE Symposium on Visual Languages and Human-Centric Computing (VL/HCC)}}. IEEE, \bibinfo{pages}{1--5}.
\newblock


\bibitem[Deibel(2005)]%
        {deibel2005team}
\bibfield{author}{\bibinfo{person}{Katherine Deibel}.} \bibinfo{year}{2005}\natexlab{}.
\newblock \showarticletitle{Team formation methods for increasing interaction during in-class group work}. In \bibinfo{booktitle}{\emph{Proceedings of the 10th annual SIGCSE conference on Innovation and technology in computer science education}}. \bibinfo{pages}{291--295}.
\newblock


\bibitem[Falkner and Munro(2009)]%
        {falkner2009easing}
\bibfield{author}{\bibinfo{person}{Katrina Falkner} {and} \bibinfo{person}{David~S Munro}.} \bibinfo{year}{2009}\natexlab{}.
\newblock \showarticletitle{Easing the transition: a collaborative learning approach}. In \bibinfo{booktitle}{\emph{Proceedings of the Eleventh Australasian Conference on Computing Education-Volume 95}}. \bibinfo{pages}{65--74}.
\newblock


\bibitem[Gro{\ss}e and Renkl(2006)]%
        {grosse2006effects}
\bibfield{author}{\bibinfo{person}{Cornelia~S Gro{\ss}e} {and} \bibinfo{person}{Alexander Renkl}.} \bibinfo{year}{2006}\natexlab{}.
\newblock \showarticletitle{Effects of multiple solution methods in mathematics learning}.
\newblock \bibinfo{journal}{\emph{Learning and Instruction}} \bibinfo{volume}{16}, \bibinfo{number}{2} (\bibinfo{year}{2006}), \bibinfo{pages}{122--138}.
\newblock


\bibitem[Henry(2013)]%
        {henry2013forming}
\bibfield{author}{\bibinfo{person}{Tyson~R Henry}.} \bibinfo{year}{2013}\natexlab{}.
\newblock \showarticletitle{Forming productive student groups using a massively parallel brute-force algorithm}. In \bibinfo{booktitle}{\emph{Proceedings of the World Congress on Engineering and Computer Science}}, Vol.~\bibinfo{volume}{1}. \bibinfo{pages}{23--25}.
\newblock


\bibitem[James and Willoughby(2011)]%
        {james2011listening}
\bibfield{author}{\bibinfo{person}{Mary~C James} {and} \bibinfo{person}{Shannon Willoughby}.} \bibinfo{year}{2011}\natexlab{}.
\newblock \showarticletitle{Listening to student conversations during clicker questions: What you have not heard might surprise you!}
\newblock \bibinfo{journal}{\emph{American Journal of Physics}} \bibinfo{volume}{79}, \bibinfo{number}{1} (\bibinfo{year}{2011}), \bibinfo{pages}{123--132}.
\newblock


\bibitem[Lee et~al\mbox{.}(2017)]%
        {lee2017exploring}
\bibfield{author}{\bibinfo{person}{Sang~Won Lee}, \bibinfo{person}{Yan Chen}, \bibinfo{person}{Noah Klugman}, \bibinfo{person}{Sai~R Gouravajhala}, \bibinfo{person}{Angela Chen}, {and} \bibinfo{person}{Walter~S Lasecki}.} \bibinfo{year}{2017}\natexlab{}.
\newblock \showarticletitle{Exploring coordination models for ad hoc programming teams}. In \bibinfo{booktitle}{\emph{Proceedings of the 2017 CHI Conference Extended Abstracts on Human Factors in Computing Systems}}. \bibinfo{pages}{2738--2745}.
\newblock


\bibitem[Porter et~al\mbox{.}(2013)]%
        {10.1145/2445196.2445250Halvingfailrates}
\bibfield{author}{\bibinfo{person}{Leo Porter}, \bibinfo{person}{Cynthia Bailey~Lee}, {and} \bibinfo{person}{Beth Simon}.} \bibinfo{year}{2013}\natexlab{}.
\newblock \showarticletitle{Halving fail rates using peer instruction: a study of four computer science courses}. In \bibinfo{booktitle}{\emph{Proceeding of the 44th ACM Technical Symposium on Computer Science Education}} (Denver, Colorado, USA) \emph{(\bibinfo{series}{SIGCSE '13})}. \bibinfo{publisher}{Association for Computing Machinery}, \bibinfo{address}{New York, NY, USA}, \bibinfo{pages}{177–182}.
\newblock
\showISBNx{9781450318686}
\urldef\tempurl%
\url{https://doi.org/10.1145/2445196.2445250}
\showDOI{\tempurl}


\bibitem[Rodr\'{\i}guez et~al\mbox{.}(2017)]%
        {10.1145/3017680.3017748vPairProgramming}
\bibfield{author}{\bibinfo{person}{Fernando~J. Rodr\'{\i}guez}, \bibinfo{person}{Kimberly~Michelle Price}, {and} \bibinfo{person}{Kristy~Elizabeth Boyer}.} \bibinfo{year}{2017}\natexlab{}.
\newblock \showarticletitle{Exploring the Pair Programming Process: Characteristics of Effective Collaboration}. In \bibinfo{booktitle}{\emph{Proceedings of the 2017 ACM SIGCSE Technical Symposium on Computer Science Education}} (Seattle, Washington, USA) \emph{(\bibinfo{series}{SIGCSE '17})}. \bibinfo{publisher}{Association for Computing Machinery}, \bibinfo{address}{New York, NY, USA}, \bibinfo{pages}{507–512}.
\newblock
\showISBNx{9781450346986}
\urldef\tempurl%
\url{https://doi.org/10.1145/3017680.3017748}
\showDOI{\tempurl}


\bibitem[Sabin and Sabin(1994)]%
        {sabin1994collaborative}
\bibfield{author}{\bibinfo{person}{Roberta~Evans Sabin} {and} \bibinfo{person}{Edward~P Sabin}.} \bibinfo{year}{1994}\natexlab{}.
\newblock \showarticletitle{Collaborative learning in an introductory computer science course}. In \bibinfo{booktitle}{\emph{Proceedings of the twenty-fifth SIGCSE symposium on Computer science education}}. \bibinfo{pages}{304--308}.
\newblock


\bibitem[Serrano-C{\'a}mara et~al\mbox{.}(2014)]%
        {serrano2014evaluation}
\bibfield{author}{\bibinfo{person}{Luis~Miguel Serrano-C{\'a}mara}, \bibinfo{person}{Maximiliano Paredes-Velasco}, \bibinfo{person}{Carlos-Mar{\'\i}a Alcover}, {and} \bibinfo{person}{J~{\'A}ngel Velazquez-Iturbide}.} \bibinfo{year}{2014}\natexlab{}.
\newblock \showarticletitle{An evaluation of students’ motivation in computer-supported collaborative learning of programming concepts}.
\newblock \bibinfo{journal}{\emph{Computers in human behavior}}  \bibinfo{volume}{31} (\bibinfo{year}{2014}), \bibinfo{pages}{499--508}.
\newblock


\bibitem[Tang et~al\mbox{.}(2023)]%
        {Tang2023VizPI}
\bibfield{author}{\bibinfo{person}{Xiaohang Tang}, \bibinfo{person}{Xi Chen}, \bibinfo{person}{Sam Wong}, {and} \bibinfo{person}{Yan Chen}.} \bibinfo{year}{2023}\natexlab{}.
\newblock \showarticletitle{VizPI: A Real-Time Visualization Tool for Enhancing Peer Instruction in Large-Scale Programming Lectures}. In \bibinfo{booktitle}{\emph{Adjunct Proceedings of the 36th Annual ACM Symposium on User Interface Software and Technology}} (San Francisco, CA, USA) \emph{(\bibinfo{series}{UIST '23 Adjunct})}. \bibinfo{publisher}{Association for Computing Machinery}, \bibinfo{address}{New York, NY, USA}, Article \bibinfo{articleno}{17}, \bibinfo{numpages}{3}~pages.
\newblock
\showISBNx{9798400700965}
\urldef\tempurl%
\url{https://doi.org/10.1145/3586182.3616632}
\showDOI{\tempurl}


\bibitem[Tang et~al\mbox{.}(2024)]%
        {tang2024vizgroup}
\bibfield{author}{\bibinfo{person}{Xiaohang Tang}, \bibinfo{person}{Sam Wong}, \bibinfo{person}{Kevin Pu}, \bibinfo{person}{Xi Chen}, \bibinfo{person}{Yalong Yang}, {and} \bibinfo{person}{Yan Chen}.} \bibinfo{year}{2024}\natexlab{}.
\newblock \showarticletitle{VizGroup: An AI-Assisted Event-Driven System for Real-Time Collaborative Programming Learning Analytics}.
\newblock \bibinfo{journal}{\emph{arXiv preprint arXiv:2404.08743}} (\bibinfo{year}{2024}).
\newblock


\bibitem[Taylor et~al\mbox{.}(2018)]%
        {taylor2018multi}
\bibfield{author}{\bibinfo{person}{Cynthia Taylor}, \bibinfo{person}{Jaime Spacco}, \bibinfo{person}{David~P Bunde}, \bibinfo{person}{Andrew Petersen}, \bibinfo{person}{Soohyun~Nam Liao}, {and} \bibinfo{person}{Leo Porter}.} \bibinfo{year}{2018}\natexlab{}.
\newblock \showarticletitle{A multi-institution exploration of peer instruction in practice}. In \bibinfo{booktitle}{\emph{Proceedings of the 23rd annual ACM conference on innovation and technology in computer science education}}. \bibinfo{pages}{308--313}.
\newblock


\bibitem[Tran et~al\mbox{.}(2004)]%
        {tran2004peer}
\bibfield{author}{\bibinfo{person}{Duc~A Tran}, \bibinfo{person}{Kien~A Hua}, {and} \bibinfo{person}{Tai~T Do}.} \bibinfo{year}{2004}\natexlab{}.
\newblock \showarticletitle{A peer-to-peer architecture for media streaming}.
\newblock \bibinfo{journal}{\emph{IEEE journal on Selected Areas in Communications}} \bibinfo{volume}{22}, \bibinfo{number}{1} (\bibinfo{year}{2004}), \bibinfo{pages}{121--133}.
\newblock


\bibitem[Vygotsky and Cole(1978)]%
        {vygotsky1978mind}
\bibfield{author}{\bibinfo{person}{Lev~Semenovich Vygotsky} {and} \bibinfo{person}{Michael Cole}.} \bibinfo{year}{1978}\natexlab{}.
\newblock \bibinfo{booktitle}{\emph{Mind in society: Development of higher psychological processes}}.
\newblock \bibinfo{publisher}{Harvard university press}.
\newblock


\bibitem[Wang et~al\mbox{.}(2021)]%
        {wang2021puzzleme}
\bibfield{author}{\bibinfo{person}{April~Yi Wang}, \bibinfo{person}{Yan Chen}, \bibinfo{person}{John Joon~Young Chung}, \bibinfo{person}{Christopher Brooks}, {and} \bibinfo{person}{Steve Oney}.} \bibinfo{year}{2021}\natexlab{}.
\newblock \showarticletitle{PuzzleMe: Leveraging Peer Assessment for In-Class Programming Exercises}.
\newblock \bibinfo{journal}{\emph{Proceedings of the ACM on Human-Computer Interaction}} \bibinfo{volume}{5}, \bibinfo{number}{CSCW2} (\bibinfo{year}{2021}), \bibinfo{pages}{1--24}.
\newblock


\bibitem[Webber and Lima(2012)]%
        {webber2012evaluating}
\bibfield{author}{\bibinfo{person}{Carine~G Webber} {and} \bibinfo{person}{Maria de F{\'a}tima Webber do~Prado Lima}.} \bibinfo{year}{2012}\natexlab{}.
\newblock \showarticletitle{Evaluating automatic group formation mechanisms to promote collaborative learning--a case study}.
\newblock \bibinfo{journal}{\emph{International Journal of Learning Technology}} \bibinfo{volume}{7}, \bibinfo{number}{3} (\bibinfo{year}{2012}), \bibinfo{pages}{261--276}.
\newblock


\bibitem[Zingaro and Porter(2014)]%
        {zingaro2014peer}
\bibfield{author}{\bibinfo{person}{Daniel Zingaro} {and} \bibinfo{person}{Leo Porter}.} \bibinfo{year}{2014}\natexlab{}.
\newblock \showarticletitle{Peer instruction in computing: The value of instructor intervention}.
\newblock \bibinfo{journal}{\emph{Computers \& Education}}  \bibinfo{volume}{71} (\bibinfo{year}{2014}), \bibinfo{pages}{87--96}.
\newblock


\end{thebibliography}

\end{document}